\documentclass{pas}

\def \ppxf {{\small P}PXF}

\def\kmsmpc {km$\,$s$^{-1}$Mpc$^{-1}$}
\def \kms {km$\,$s$^{-1}$}
\def \oiii {[O{\small~III}]}

\def \sii {[S{\small~II}]}
\def \nii {[N{\small~II}]}
\def \ha {H$\alpha$}
\def \hb {H$\beta$}

\usepackage{multirow}
\usepackage{gensymb}
\usepackage{aas_macros}
\usepackage{preamble}

\begin{document}

\lefttitle{Constraining the timescale of activity in Mrk 1018}
\righttitle{R. McElroy et al.}

\jnlPage{1}{4}
\jnlDoiYr{2021}
\doival{10.1017/pasa.xxxx.xx}

\articletitt{Research Paper}

\title{Constraining the timescale of activity in Mrk 1018}

\author{Rebecca McElroy$^{1,2,3}$, Scott M. Croom$^2$, Thomas Dunn$^2$}

\affil{$^1$Centre for Astrophysics, University of Southern Queensland, Toowoomba, QLD 4350, Australia\\ $^2$Sydney Institute for Astronomy (SIfA), School of Physics, The University of Sydney, NSW 2006, Australia\\
$^3$Astrophysics and Space Technologies Research Centre, 12 Wally's Walk, Macquarie University, 2109, Australia}

\corresp{R. McElroy, Email: \href{mailto:rebecca.mcelroy@unisq.edu.au}{rebecca.mcelroy@unisq.edu.au}}


\history{(Received xx xx xxxx; revised xx xx xxxx; accepted xx xx xxxx)}

\begin{abstract}

We present analysis of an extended emission line region projected 3.77 kpc from the nucleus of the canonical changing-look AGN
Mrk~1018, found in MUSE integral-field observations from the Close AGN Reference Survey. 
Emission-line ratios show that the region exhibits AGN ionisation but has a very low velocity dispersion, $77 \pm 1$\,km\,s$^{-1}$, ruling out an ionised outflow. 
The lower limit on the  light travel time is $12{,}300 \pm 2{,}300$\,yr, implying that 
the nucleus has therefore been active for at least this long.
This is much longer than the duration of either observed
changing-look transition, which occurred over several years. This implies that Mrk 1018's multi-year variations are occurring within a longer term active phase. 

\end{abstract}


\maketitle

\section{Introduction}

`Changing Look' active galactic nuclei (CLAGN) are a highly variable subclass of active galaxies. Not only do they vary extensively in brightness but their spectral properties also change. This has made them a popular test-bed for our understanding of how supermassive black holes (SMBH) grow and regulate. Mrk 1018 is a canonical CLAGN at z = 0.043 that has been studied extensively \citep{cohen_variability_1986, 2016A&A...593L...8M, 2016A&A...593L...9H, 2017A&A...607L...9K, saha_close_2025}. Since its first spectral observation it has transitioned twice and its long-term variability is described in \citet{2025arXiv251025156D}. 

Approximately 1400 CLAGN have been found to date \citep{2026arXiv260725839C}. Current observations suggest that they are a rare population, making up approximately 0.1--1\% of the general broad-line AGN population \citep[e.g.,][]{macleod_changing-look_2019}, though they represent up to 15--20\% of extremely variable quasars \citep{2016MNRAS.457..389M, 2026ApJ...997..100W}. Whether CLAGN represent a short-lived phase that all AGN experience during their activity cycles, or an intrinsically distinct, rare sub-population, is still debated. 

There is evidence that CLAGN represent a transient phase common to all accreting supermassive black holes (SMBHs). Active galactic nuclei are known to undergo long-term, episodic accretion cycles (or ``flickering'') on timescales of $10^4$--$10^5\,\mathrm{yr}$ \citep[e.g.,][]{2012ApJ...747L..33E, 2015MNRAS.451.2517S, cars_dr1}. Rather than requiring unique host galaxy or black hole properties, the changing-look phenomenon appears to occur whenever a system's accretion rate crosses a universal critical threshold ($\lambda_{\mathrm{Edd}} \equiv L_{\mathrm{bol}}/L_{\mathrm{Edd}} \sim 0.01$--$0.02$) \citep{noda_explaining_2018, 2026ApJ..1007L...4J}. Because every growing SMBH will experience accretion rate changes during its multi-Gyr growth history, any active galaxy would temporarily manifest as a CLAGN as its accretion transitions across this boundary.

\begin{figure}
    \centering
    \includegraphics[width=0.7\linewidth]{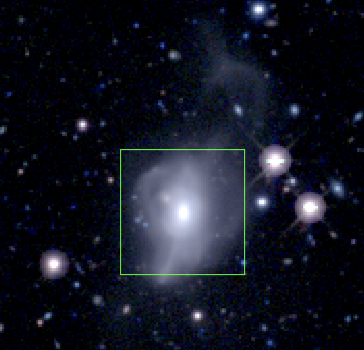}
    \caption{A three colour DESI Legacy image of Mrk 1018 with the MUSE IFU footprint overlaid. Clear tidal features are seen across the galaxy and beyond the FoV. }
    \label{fig:legacy}
\end{figure}

Physically, the CLAGN phase corresponds to a global accretion state transition, analogous to the soft/hard state transitions observed in black hole X-ray binaries \citep{noda_explaining_2018, ricci_changing-look_2023}. When the local mass accretion rate drops below $\lambda_{\mathrm{Edd}} \approx 0.01$, the inner, radiatively efficient, geometrically thin accretion disk and warm corona collapse or truncate into an optically thin, advection-dominated accretion flow (ADAF) \citep{2019ApJ...883...76R,  saha_close_2025}. This sudden drop in ionizing UV continuum radiation causes the disappearance of the broad emission lines on timescales of months to a few years ($t_{\mathrm{fp}} \sim \mathrm{years}$) \citep{2025A&A...693A..35J, ricci_changing-look_2023}. Conversely, an influx of gas pushing the accretion rate back above $\lambda_{\mathrm{Edd}} \sim 0.01$ rebuilds the thin disk and re-illuminates the broad line region (BLR) (a ``turn-on'' event). In this picture, the CLAGN state is not a distinct class of AGN, but a phase during the lifetime of an accretion flow.

\begin{figure}
    \centering
    \includegraphics[width=0.99\linewidth]{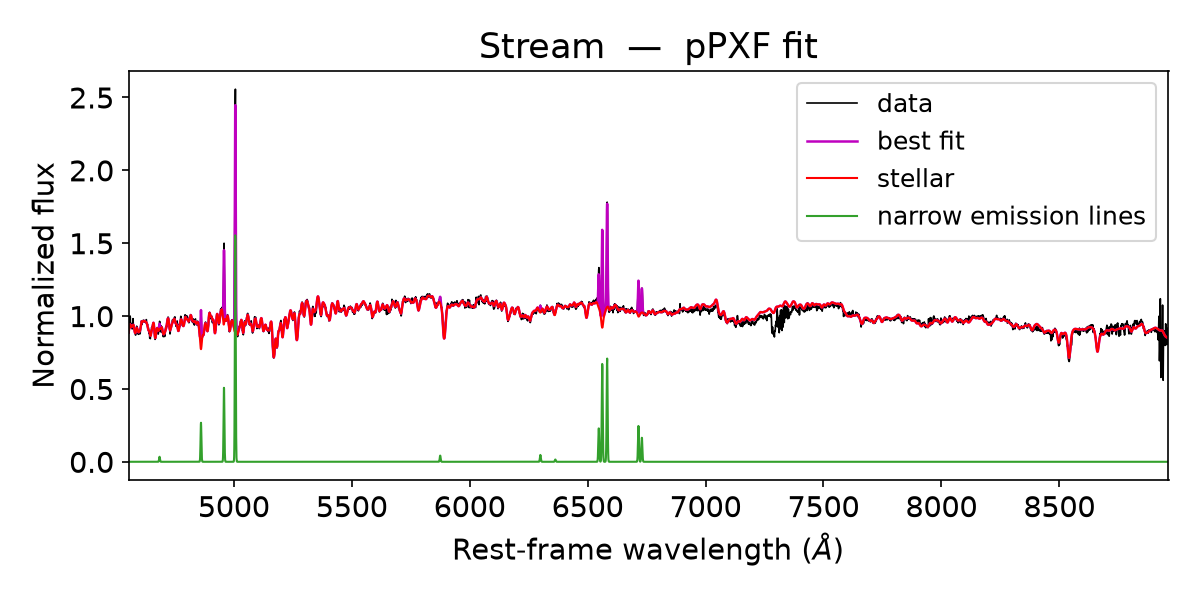}
    \caption{Example of the {\sc ppxf} fit to the stream spectrum. Data is shown in black, stellar fit in red, best fit in pink, and the narrow line fit in green.}
    \label{fig:ppxf}
\end{figure}

\begin{figure*}[h]
    \centering
    \includegraphics[width=0.85\linewidth]{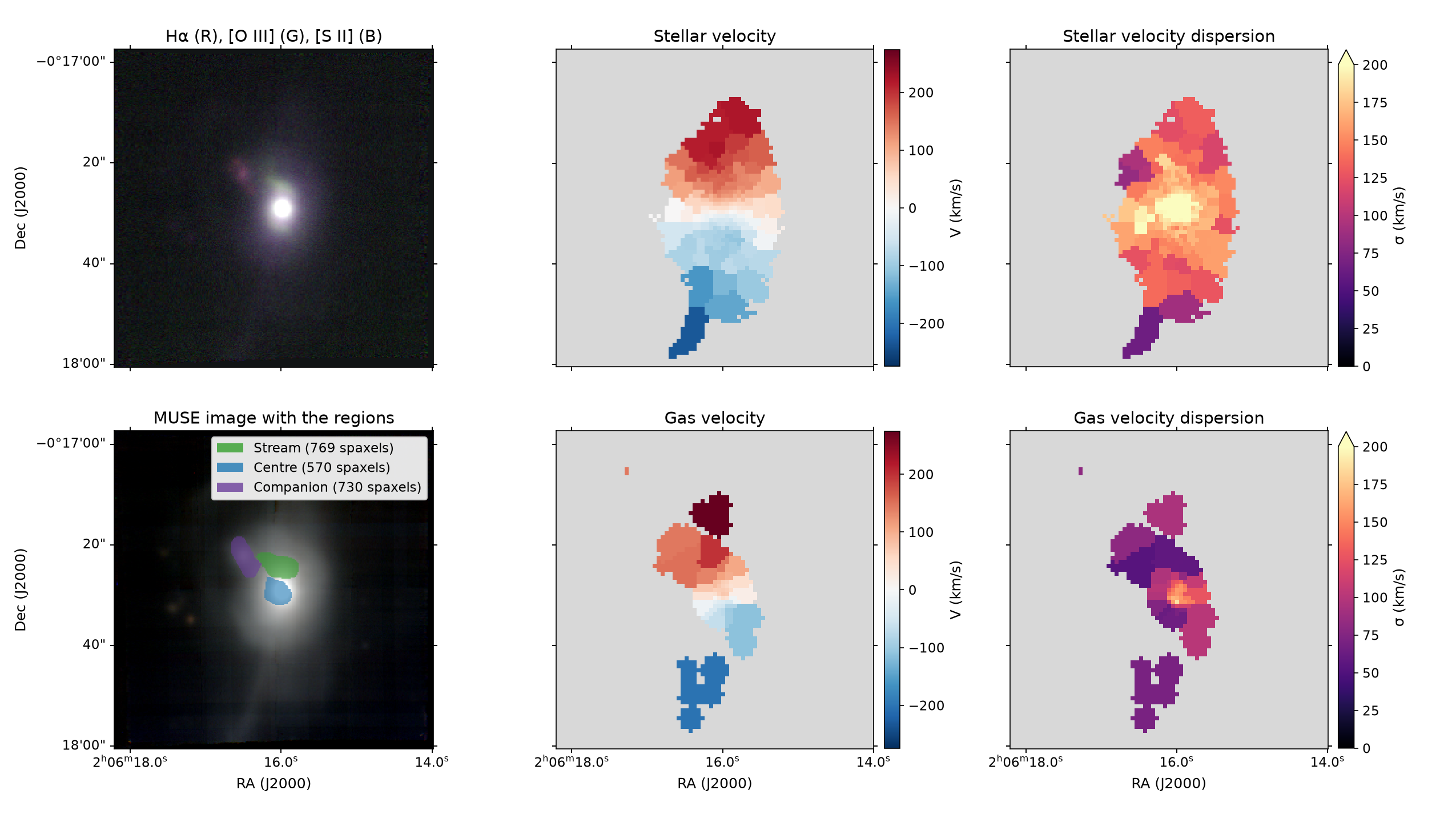}
    \caption{\textit{Top left:} RGB image with \protect{\ha} as red, [OIII] as green, and [SII] as blue. The companion galaxy appears in red, the stream in green, and the host galaxy appears white. \textit{Bottom left:} MUSE continuum image with regions superimposed. \textit{Right:} Maps of the stellar and gaseous velocity and velocity dispersion fits from {\sc ppxf}. We use Voronoi binning with a target S/N of 20 on the \ha\ and continuum.}
    \label{fig:cube}
\end{figure*}

While light curves capture the rapid ($\sim\mathrm{years}$) transitions, they are limited to the late 19th century onwards. This does not allow us to probe the ``flickering" timescales discussed above. If we want to measure how long a given AGN has been switched on, we need to look beyond the nucleus to larger spatial scales where the light travel time creates a physical illumination record of the central engine's output. Using extended emission-line regions (EELRs) \citep{2009MNRAS.399..129L, 2012MNRAS.420..878K, 2024MNRAS.530.1624K} we can infer the history of the ionizing continuum, placing constraints on the exact length of time an AGN has been switched on.

In this paper we describe an EELR found in Mrk 1018 and use it to place a constraint on the duration of the AGN's active phase. In Section \ref{sec:data} we describe the data we analysed.  In Section \ref{sec:analysis} we describe the data analysis. Our results are presented in Section \ref{sec:discussion} prior to the conclusion in Section \ref{sec:conc}. We assume a cosmology with $\Omega_{\rm m}=0.3097$, $\Omega_\Lambda=0.6894$ and  $H_0=67.66$\,\kmsmpc \citep{2020A&A...641A...6P}.

\section{Data}\label{sec:data}

We observed Mrk 1018 with the Multi-Unit Spectroscopic Explorer (MUSE, \citet{2010SPIE.7735E..08B}) as part of the Close AGN Reference Survey \citep{cars_dr1, 2022Msngr.187....3M}. This led to the discovery that it had changed type for a second time \citep{2016A&A...593L...8M}.

Mrk 1018 is a merger remnant. It has clear tidal streams shown in Figure \ref{fig:legacy} and what appears to be a small infalling companion near the nucleus. Upon examination of the emission lines in the cube we found that there were small regions of heightened emission in this area. To make these features clearer we performed a 4x4 boxcar bin of the MUSE datacube to increase the signal-to-noise (S/N) in faint structures. We then isolated the key emission lines in velocity space and created an image, shown in the top left of Figure \ref{fig:cube}. Here \ha\ is shown in red, \oiii\ is shown in green, and \sii\ is shown in blue. Two highly ionised regions outside the nucleus pop out, one aligned with the infalling companion and one between the nucleus and the companion. Henceforth these regions will be defined as \textit{companion} and \textit{stream}. We created spectra for the companion, stream, and nuclear regions by averaging all spectra within a spatial mask created interactively using the \textsc{Glue} data visualisation environment \citep{2019zndo...3385920R}.

\section{Analysis}\label{sec:analysis}

We model the stellar continuum and emission lines  with \ppxf \citep{2017MNRAS.466..798C} using the E-MILES simple stellar population templates \citep{2010MNRAS.404.1639V}. We use a stellar kinematic component, a narrow emission-line component for all emission lines, and an optional broad emission-line component for \ha\ and \hb\ in the centre. We also use a 10th degree multiplicative polynomial to account for residual continuum shape differences between the data and templates with no additive polynomial. We perform this analysis on the combined spectra from the companion, stream, and centre and on the binned datacube, using Voronoi bins based on the continuum S/N for the stellar parameters and the \ha\ S/N for the emission parameters. In both cases we adopt a target S/N of 20, with the note that this does result in coarse bins -- particularly in the gas kinematics. The fit to the stream spectrum is shown in Figure \ref{fig:ppxf} and the stellar and emission properties for the whole galaxy are shown in Figure \ref{fig:cube}. We approximate the error on the measured velocity dispersion using a bootstrap of 200 refits.

These models also allow us to measure the velocity dispersion of each of the regions. We find that the companion's velocity dispersion sits at the spectral resolution limit of 50 \kms. This value is consistent with star formation and rotation, though rotation is not observed due to the binning.  The narrow gas component in the centre has a higher velocity dispersion of 148 $\pm$ 2 \kms. The stream has a velocity dispersion of 77 $\pm$ 1 \kms. We can check this against the spatially resolved velocity maps in Figure \ref{fig:cube}. While the stellar velocity dispersion remains high out to the location of the stream, the gas dispersion is below 100 \kms -- consistent with the measurement for the summed region. 

The low velocity dispersion of the stream indicates that this is not an outflow that has been ionised by shocks, as they typically have velocity dispersions of hundreds of \kms \citep{Kakkad_2022}.

We construct diagnostic (BPT) diagrams using the modelled emission line fluxes \citep{1981PASP...93....5B}. The theoretical maximum-starburst boundary of \citet{2001ApJ...556..121K} and the empirical star-forming/composite demarcation of \citet{2003MNRAS.346.1055K} were used to classify excitation in the \nii-BPT; for the \nii-BPT, the Seyfert/LINER separation of \citet{2006MNRAS.372..961K} was also applied. These BPT diagrams are shown in Figure \ref{fig:bpt} where each region is represented by a different coloured point. The companion lies in the star-forming region, while the stream and centre lie in the AGN region. Given the low dispersion, that rules out strong shocks, this tells us that the stream was ionised by radiation from the AGN. 

We can measure the \sii\ ratio using our emission line fits. The ratio for the stream, $R = 1.49 \pm 0.13$, sits at the low-density limit at $T_e = 10^{4}$\,K, so we can only infer an upper limit of $n_e < 137\,\mathrm{cm^{-3}}$. The recombination time, $t_\mathrm{rec} = 1/(n_e \alpha_B)$ with $\alpha_B = 2.59\times10^{-13}\,\mathrm{cm^{3}\,s^{-1}}$, is therefore $\gtrsim 9\times10^{2}\,\mathrm{yr}$, and longer if the gas is less dense. This is much longer than the several-year changing-look transition, so the stream does not track changes on that timescale and instead records its average output over $\gtrsim10^{3}\,\mathrm{yr}$.

We can measure the projected distance between the nucleus, taken to be the brightest pixel, and the centre of the stream. While this may not be the true distance, it is a lower limit. We measure the projected distance to be $4.3\pm0.8$", taking an error of one binned pixel. At the redshift of the galaxy this is 3.77 kpc. This equates to a light travel time of 12,300$\pm 2300$ years.

\section{Discussion}\label{sec:discussion}

\begin{figure}
    \centering
    \includegraphics[width=\linewidth]{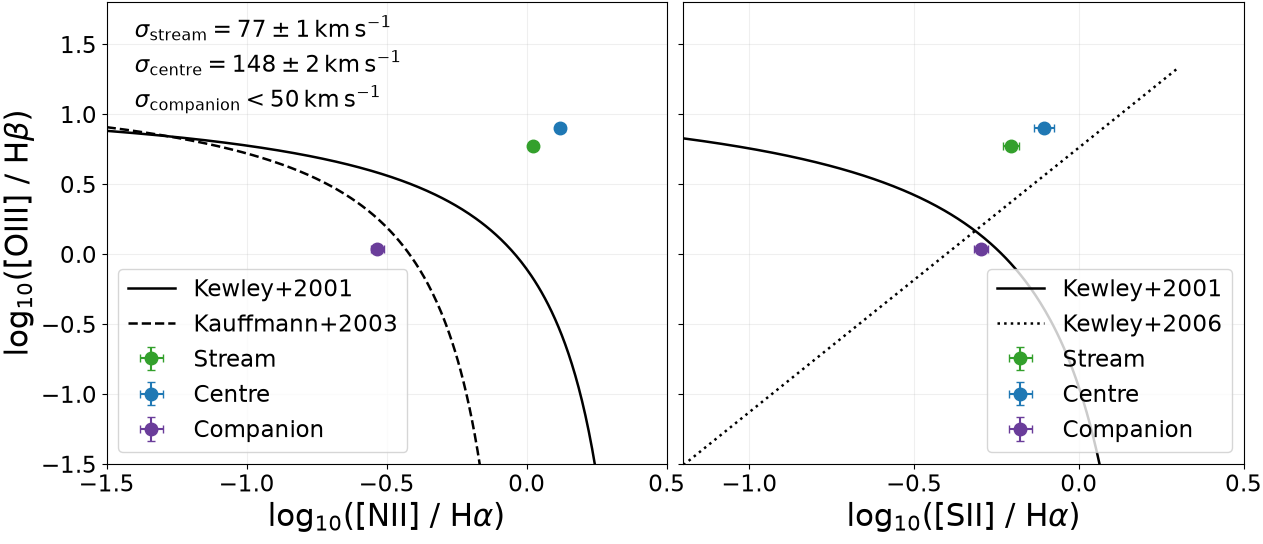}
    \caption{BPT diagrams for the three region spectra. Demarcation lines are used to separate the AGN, composite, and star forming regions. Centre is in blue, companion in purple, and stream in green. The centre and stream fall into the AGN region of the diagrams while the companion is in the star forming region.}
    \label{fig:bpt}
\end{figure}


The stream occupies the AGN region of both the \nii\ and
\sii\ BPT diagrams (Figure~\ref{fig:bpt}), while the companion
lies in the star-forming region, demonstrating that the two structures have different primary ionisation 
sources even though they are close together on the sky.
The stream velocity dispersion of 77 \kms is much too low to be the result of shock-ionised
outflows from the nucleus \citep{Kakkad_2022}. This leaves photoionisation by the nucleus as the remaining explanation. 

The projected light travel time of 12,300 $\pm$ 2,300 yr is a lower limit on the true
nucleus to stream light travel time. \citet{kim_recoiling_2018} measure an
inclination of 49\degree\ for Mrk~1018 using {\sc GALFIT}. Using this for the plane in which
the stream lies, the deprojected separation falls between $3.77$ and
$5.75$\,kpc, depending on the orientation of the nucleus--stream vector
relative to the line of nodes, corresponding to a light travel time of
$12{,}300$--$18{,}700$\,yr. We will continue to quote the projected value as a conservative
lower limit throughout, since the inclination measured for Mrk 1018
has a large uncertainty as it is a merger remnant.

Extended ionised gas of this kind is often found in interacting and
merging systems. In the TELPERION survey, \citet{2024MNRAS.530.1624K} find distant EELRs in 2--5\% of AGN depending on sample selection. This rises
to $12 \pm 6$\% among interacting and merging AGN. In a previous paper, they 
found that kinematically quiescent EELRs are almost always photoionised
tidal debris, seen in roughly 10\% of their galaxies with tidal tails
\citep{2022MNRAS.510.4608K}. The best-studied example of this, Hanny's Voorwerp near IC~2497, is
 ionised gas in a tidal structure
\citep{2009MNRAS.399..129L,2012MNRAS.420..878K}. It is perhaps not surprising that a galaxy like Mrk~1018 has an EELR. Its stellar
populations are old and it contains little ionised gas outside the nuclear
region, so it is likely that gas within several kpc of the centre -- where it can be illuminated -- has been supplied by the merger. This means that in population studies of CLAGN, we would expect to find more EELRs in merging systems.

We found the recombination time of the stream to be $\gtrsim 900$\,yr. This means that the ionisation state of the stream can persist for
approximately 7\% of the light travel time, and the observed line ratios
constrain an integrated illumination history on this timescale. This means that the gas
must have been ionised by the nucleus at some point within $\sim$900\,yr of
the light travel time. However, it does mean that a decline in nuclear
output on this timescale would not necessarily be registered in these
ratios.

Mrk~1018 has undergone several changing-look transitions
\citep{cohen_variability_1986,2016A&A...593L...8M}, and its photometric record extends over
roughly a century \citep{2025arXiv251025156D}. The timescale we  derive here,
$\gtrsim$12{,}300\,yr, is three orders of magnitude longer, and implies
that the nucleus has been active far longer than the duration of either
observed transition. Mrk 1018's changing-look events  therefore appear to be rapid variation
within a longer active phase rather than the onset or termination of
activity.

\section{Conclusions}
\label{sec:conc}

In this work we have used MUSE IFU observations from the Close AGN Reference Survey to
identify AGN-ionised gas in a tidal stream $\sim$4" from the nucleus
of Mrk~1018, and used its light travel time to constrain a lower limit on how long the AGN
has been active. We find:

\begin{itemize}
\item The stream and the nucleus are AGN ionised, while the nearby
companion is dominated by star formation, showing that the two regions
have different primary ionisation sources.

\item The stream velocity dispersion of $77 \pm 1$\,\kms\ is much lower than shock-ionised outflows, making photoionisation from the nucleus and a resulting EELR likely.

\item The lower limit on the projected light travel time between the nucleus and the stream is
$12{,}300 \pm 2{,}300$\,yr -- three
orders of magnitude longer than the observed changing look transitions. 

\end{itemize}

This implies that Mrk 1018's changing-look events are modulations within a
sustained active phase rather than the signatures close to the beginning or end of an AGN phase. Since Mrk 1018 is a single object, we cannot make any statement about the properties of CLAGN as a whole. Applying this
approach to the ever-increasing sample of CLAGN with integral-field
observations could allow us to go from  individual constraints to a distribution of
activity timescales, providing a direct observational test of the
$10^{4}$--$10^{5}$\,yr flickering timescales inferred statistically \citep{2015MNRAS.451.2517S}.

\section*{Data Availability}

This paper uses MUSE data from the Close AGN Reference Survey in programme 094.B-0345 (PI: Husemann). It is publicly available through the \href{https://cars.aip.de/}{CARS Data Release} and the ESO archive.

\bibliographystyle{paslike}
\bibliography{references}

\end{document}